# Hydrogenation and Hydro-Carbonation and Etching of Single-Walled Carbon Nanotubes

Guangyu Zhang, Pengfei Qi, Xinran Wang, Yuerui Lu, David Mann, Xiaolin Li and Hongjie Dai*

*Department of Chemistry and Laboratory for Advanced Materials, Stanford University, Stanford, CA 94305*

RECEIVED DATE (automatically inserted by publisher); E-mail: hdai@stanford.edu

Understanding the interactions between molecules and single-walled carbon nanotubes (SWNTs) is of fundamental and practical importance.[1] While weak adsorption of molecular hydrogen on SWNTs has been widely investigated and debated, covalent reactions between hydrogen and SWNTs are less explored. Only a handful experiments have been done at room temperature (RT) thus far with a main finding that SWNTs are hydrogenated by atomic hydrogen or hydrogen plasma with an atomic coverage up to ~65%.[2-5] It is also found recently that hydrogenation plays a role in affecting the growth of SWNTs in plasma-enhanced chemical vapor deposition.[6]

Here we present a systematic experimental investigation of the reactions between H-plasma and SWNTs at various temperatures. We reveal structural, infrared (IR), Raman spectroscopic and electrical properties of hydrogenated SWNTs. Further, we uncover hydrogen-plasma cutting and -etching of SWNTs and the nanotube diameter dependence of the etching effect.

We used a quasi remote radio-frequency (RF) plasma system to generate H-plasma under a $H_2$ gas flow at 1 torr and a RF power of 30W. Chemical vapor deposition (CVD)[1,7] grown SWNTs on substrates were placed downstream of the plasma source and exposed to the H-radicals carried over from the plasma by the gas flow (see Supplementary Information).

Systematic atomic force microscopy (AFM) imaging reveals an apparent increase in the topographic heights of nanotubes (Fig.1) after H-plasma treatment at RT for 3 minutes. The height increase is ~3Å±1Å based on measurements (Fig.1) over 100 nanotubes before and after exposure to H-plasma. We attribute this height increase to hydrogenation of the sidewalls of SWNTs. A covalently bond H to the nanotube sidewall adds ~ 1Å. Also, calculations have shown that hydrogenation of a SWNT lead to deformation and relaxation in the carbon network.[8] These are the likely causes to the observed nanotube size increase. Note that the height change appears uniform along the lengths of nanotubes, indicating dense hydrogenation sites (spacing of sites should be well below the AFM lateral resolution of ~5-10nm). This is the first time that SWNT 'swelling' is observed due to hydrogenation.

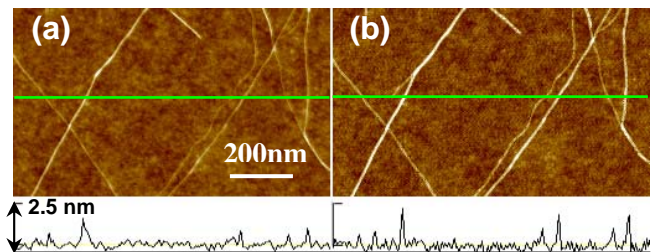

**Fig.1** AFM images of SWNTs a) before and b) after RT plasma treatment. Bottom: topography height profiles along the green lines in the images.

To further elucidate hydrogenation of SWNTs, we have carried out IR spectroscopy measurements of dense vertically aligned SWNT films[6] before and after RT H-plasma treatment and subsequent step-wise thermal annealing up to 500ºC. After RT H-plasma treatment of the nanotubes, vibration bands in the range of 2750-3000cm$^{-1}$ with peaks centered at ~2850cm$^{-1}$ and ~2920cm$^{-1}$ are observed (Fig.2a). The band centered at 2920cm$^{-1}$ can be assigned to $sp^3$ CH stretching or asymmetric stretching of $sp^3$ $CH_2$ groups.[9] The 2850cm$^{-1}$ peak matches symmetric stretching of $sp^3$ $CH_2$. These results suggest the existence of $sp^3$ CH on the sidewall of the SWNTs as well as $sp^3$ $CH_2$ species likely at the defect sites and ends of SWNTs. Note that abundant ends exist at the top of our vertically aligned SWNTs (see Supplementary Information) used for the IR experiments. Upon thermal annealing, we observed that dehydrogenation mainly occurs between 400-500ºC (Fig.2a), suggesting the reversal of C-$H_x$ formation at ~500ºC likely via the release of hydrogen.

Raman spectroscopy measurements reveal that after RT H-plasma treatment, the ratio between the integrated area of disordered peak (D peak) and graphitic peak (G peak) (D/G) increases from 16% to 91% (Fig.2b). Hydrogenation causes $sp^3$ C-$H_x$ formation, introduces disorder on SWNT sidewalls and thus increases the D band intensity. Upon annealing at 500ºC, the D/G ratio reduced to 19% (Fig.2b), suggesting that dehydrogenation leads to nearly full recovery of the structures of the SWNTs.

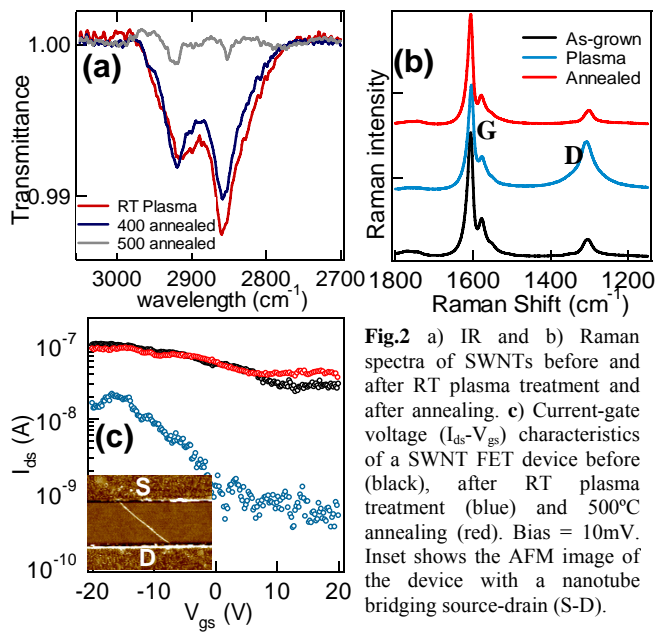

**Fig.2** a) IR and b) Raman spectra of SWNTs before and after RT plasma treatment and after annealing. c) Current-gate voltage ($I_{ds}$-$V_{gs}$) characteristics of a SWNT FET device before (black), after RT plasma treatment (blue) and 500ºC annealing (red). Bias = 10mV. Inset shows the AFM image of the device with a nanotube bridging source-drain (S-D).

Next, we carried out electrical transport measurements of individual SWNTs (see Supplementary Information) to glean the effect of hydrogenation to the electrical properties. With over 100 single tube devices investigated, we consistently observed drastic conductance decreases after RT H-plasma treatment and nearly full recovery of the conductance after 500ºC annealing (Fig.2c). The conductance decrease upon hydrogenation can be attributed to the $sp^2$-$sp^3$ structure change of SWNT, leading to localization of π-electrons. It was suggested that hydrogenation could open up or increase the bandgap of SWNTs.[8] Accompanied by large decreases in the overall conductance, we indeed observed that the conductance of metallic and quasi-metallic SWNTs exhibited stronger gate-voltage dependence upon hydrogenation (Fig.2c). This was likely due to the more semiconducting nature of hydrogenated tubes.

The microscopy, spectroscopy and electrical data above suggests that under the specific RT remote-plasma condition (30W for 3 min), hydrogenation of SWNTs leads to structural and property changes that are largely reversible via dehydrogenation at 500 ºC. However, under harsher plasma conditions at RT (i.e., higher power, such as 50W and longer plasma time up to 10 min), we observed irreversible etching and even complete removal of SWNTs especially for those with smaller diameters imaged by AFM (Fig.3a). Increasing the sample temperature from RT up to 400ºC during the 30W/3min plasma exposure also led to cutting and etching of SWNTs (Fig. 3b-3d). These findings suggest that denser H-plasma and higher reaction temperatures can etch and completely remove SWNTs likely via gasification of nanotubes to hydrocarbon molecules (hydro-carbonation).

Our AFM results (Fig.3) combined with IR data (Supplementary Info) reveal that for high temperature H-plasma treatment of SWNTs up to 400ºC (below the de-hydrogenation T of ~ 500ºC), H-etching and cutting is accompanied by hydrogenation of SWNTs with the existence of C-$H_x$ species on the nanotubes (Fig.4b). These species vanish upon thermal annealing at 500ºC (Supplementary Info), leaving the nanotubes in a cut and de-hydrogenated form.

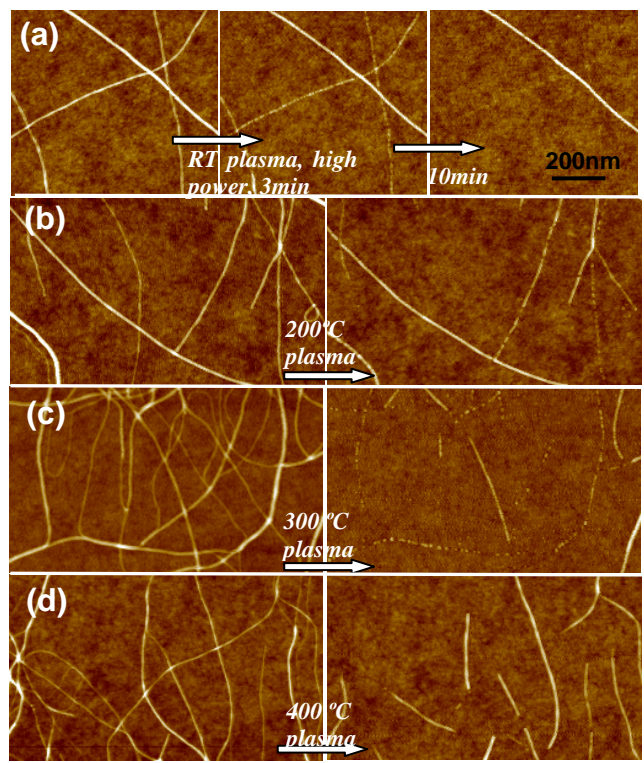

**Fig.3** (a)–(d) AFM images of four SWNT samples before and after H-plasma treatment under various conditions. For (a), RF power is 50W. For (b)-(d), RF power is 30W and H-plasma is 3 min. (e) Diameter of the smallest tubes that 'survived' H-plasma (30W, 3min) at various T.

A noteworthy phenomenon observed during high temperature H-plasma treatments of SWNT is that smaller diameter nanotubes are H-etched and cut more easily than larger tubes. We observed complete removal of SWNTs with diameters ≤ ~1nm by H-plasma treatment at 200 to 400ºC (Fig.3b-3d). If we define "surviving" SWNTs as those with no visible cuts over ≥ ~500nm length under AFM, the diameters of the smallest "surviving" SWNTs exhibit a clear increasing trend under higher treatment temperatures (Fig3e). That is, only larger SWNTs can survive higher temperature H-treatment. These results suggest higher reactivity of SWNTs with smaller diameters towards hydro-carbonation, due to the higher curvature and strain in the structures. Calculations have predicted that smaller diameter SWNTs should exhibit higher hydrogenation reactivity due to higher curvature,[10] but no theoretical account exists for hydrogen-etching of SWNTs.

Dense and small cuts are typically observed along SWNTs with diameter in the range of ~1 to 1.5nm after H-treatment at 200ºC and 300ºC (Fig.3b, 3c). This result suggests that relatively small nanotubes are attacked and cut non-discriminately at dense sites along their lengths. These tubes are completely removed when H-plasma treated at ≥ 400ºC. For larger nanotubes (diameter >~1.7nm), treatment by H-plasma at ≥ 400ºC causes fewer and wider cuts (Fig.3d), suggesting that H-etching of large nanotubes at high temperature likely starts at defect sites (or strained curved sites due to pinning interaction of the substrate) and then propagates along the tube length.

In summary, we have revealed structural enlargement, drastically reduced electrical conductance and increased semiconducting nature of SWNTs upon sidewall hydrogenation (Fig.4a) at room temperature. These changes are reversible upon thermal annealing at 500ºC via dehydrogenation. Harsh plasma or high temperature reactions lead to etching of nanotube (Fig.4b), and smaller SWNTs are markedly less stable against hydro-carbonation than larger tubes. Our results are fundamental and may have implications to basic and practical applications including hydrogen storage, sensing, band-gap engineering for novel electronics and new methods of manipulation, functionalization and etching and selection of nanotubes based on diameter.

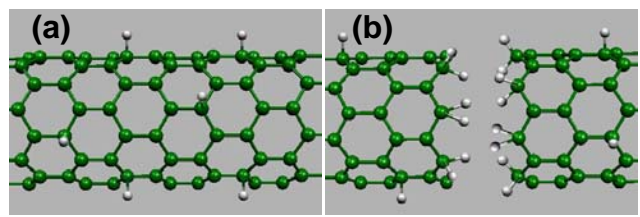

**Fig. 4** Schematic drawing of **a**) hydrogenation and **b**) hydrogenation plus etching of SWNT by H plasma at elevated temperatures (200-400ºC) or harsh plasma conditions at room temperature.

**Acknowledgement.** This work was partially supported by Stanford GCEP.

**Supporting Information Available:** Experimental details are available free of charge via the Internet at http://pubs.acs.org.

TOC Entry

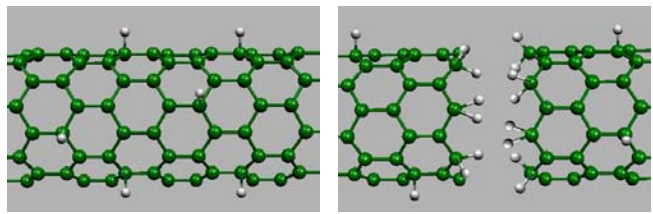

## Abstract

We present a systematic experimental investigation of the reactions between hydrogen plasma and single-walled carbon nanotubes (SWNTs) at various temperatures. Microscopy, infrared (IR) and Raman spectroscopy and electrical transport measurements are carried out to investigate the properties of SWNTs after hydrogenation. Structural deformations, drastically reduced electrical conductance and increased semiconducting nature of SWNTs upon sidewall hydrogenation are observed. These changes are reversible upon thermal annealing at 500ºC via dehydrogenation. Harsh plasma or high temperature reactions lead to etching of nanotube likely via hydro-carbonation. Smaller SWNTs are markedly less stable against hydro-carbonation than larger tubes. The results are fundamental and may have implications to basic and practical applications including hydrogen storage, sensing, band-gap engineering for novel electronics and new methods of manipulation, functionalization and etching of nanotubes.